\documentclass[a4paper]{article}[12]

\usepackage[pages=all, color=black, position={current page.south}, placement=bottom, scale=1, opacity=1, vshift=5mm]{background}
\SetBgContents{
} % copyright

\usepackage[margin=1in]{geometry} % full-width
\usepackage{graphicx,subcaption,lipsum}
\usepackage{amsmath}
\usepackage{amsthm}
\usepackage{amssymb}
\usepackage[english]{babel}
\addto{\extrasenglish}{%
  
}
\usepackage{subcaption}
\usepackage{booktabs}
\usepackage{listings}
\usepackage{multirow}

\usepackage[utf8]{inputenc}
\usepackage{hyperref}
\hypersetup{
	unicode,
	colorlinks,
	breaklinks,
	urlcolor=blue, 
	linkcolor=blue, 
	pdfauthor={Author One, Author Two, Author Three},
	pdftitle={A simple article template},
	pdfsubject={A simple article template},
	pdfkeywords={article, template, simple},
	pdfproducer={LaTeX},
	pdfcreator={pdflatex}
}

\usepackage[numbers]{natbib}
\usepackage[nottoc]{tocbibind}

\usepackage{graphicx}
\theoremstyle{plain}

\theoremstyle{definition}

\usepackage{graphicx, color}
\graphicspath{{fig/}}
\usepackage{algorithm, algpseudocode} % use algorithm and algorithmicx for typesetting algo
\usepackage{mathrsfs} % for \mathscr command
\title{Log-Gaussian Cox Processes for Spatiotemporal Traffic Fatality Estimation in Addis Ababa}

\author{Yassin Tesfaw Abebe$^{1,3}$ \and Abdu Mohammed Seid$^1$ \and Lassi Roininen$^2$}

\date{
	$^1$Bahir Dar University, Ethiopia\\%\texttt{abdum442@yahoo.com}\\%
	$^2$Lappenranta-Lahti University of Technology, Finland \\ %\texttt{Lassi.Roindinen@lut.fin}\\
	$^3$Mekdela Amba University, Ethiopia \\  %\texttt{ytesfaw@yahoo.com}\\[2ex]%
}

\begin{document}
%\pagecolor{yellow!15}
\maketitle
	
\begin{abstract}

We investigate the spatiotemporal dynamics of traffic accidents in Addis Ababa, Ethiopia, using 2016 – 2019 data. We formulate the traffic accident intensity as a log-Gaussian Cox Process and model it as a spatiotemporal point process with and without fixed and random effect components that incorporate possible covariates and spatial correlation information. The covariate includes population density and distance of accident locations from schools, from markets, from bus stops and from worship places. We estimate the posterior of the state variables using integrated nested Laplace approximations with stochastic partial differential equations approach by considering Mat\'ern prior. Deviance and Watanabe - Akaike information criteria are used to check the performance of the models. We implement the methodology to map traffic accident intensity over Addis Ababa entirely and on its road networks and visualize the potential traffic accident hotspot areas. The comparison of the observation with the model output reveals that the covariates considered has significant effect for the accident intensity. Moreover, the information criteria results reveal the model with covariate performs well compared with the model without covariates. We obtained temporal correlation of the log-intensity as 0.78 indicating the existence of similar traffic fatality trend in space during the study period.

\vspace{.25in}
\noindent\textbf{Keywords:} Hierarchical Bayesian modeling, Poisson point process, R-INLA, SPDE approach, traffic accident, Ethiopia.
\end{abstract}

\section{Introduction} \label{sec:intro}
Road traffic accident is a global social and public health challenge.
According to the Global Status Report on Road Safety \citep{world2019global}, by the World Health Organization, about 1.35 million people die every year in car accidents with another 20 -- 50 million injuries that do not result in death.
As of 2019, the WHO has identified road traffic accidents as one of the main causes of death for people of all ages and the eighth leading cause of disability and death globally. 
Road traffic accidents are consequences of the increased mobility of today’s society. 
The WHO report indicated that road traffic injuries are the primary cause of death for children, with young adults aged 5–29 years being the most vulnerable. Furthermore, low- and middle-income countries account for 92\% of global road fatalities, despite having approximately 60\% of the world's vehicles. 
Traffic accidents in Ethiopia have increased during the past five years. The number of road traffic accidents countrywide increased from 157,326 in 2014–2015 to 183,472 in 2018–2019 \citep{berhanu2023spatial}. 
This represents a 16.6\% increase in the last five years. 
Governments worldwide announced a Second Decade of Action for Road Safety 2021–2030 in response to similar issues in numerous countries, with a goal of reducing traffic fatalities and injuries by at least 50\%  \citep{world2021global}. 
The number of traffic incidents in Addis Ababa, Ethiopia rose by 37.9\% from 4,262 in 2014/2015 to 5,874 in 2018/2019. 
Road traffic crashes in Ethiopia cause 13 deaths and 37 injuries daily \citep{berhanu2023spatial}, with Addis Ababa accounting for 10\% of the deaths and 26\% of the injuries despite having 56\% of the registered vehicles. 
These data explicitly demonstrate the need to identify factors and predict traffic accident intensities to devise strategies for reducing the number of accidents.

Mathematical and statistical methodologies play a crucial role in the analysis and modeling of traffic accidents using data. 
Researchers have developed various techniques for detecting areas with high accident concentration, known as "hot spots", which can help identify areas in need of increased attention and resources \citep{hashimoto2016development, shafabakhsh2017gis}. 
In the early phases of traffic accident investigations, researchers used spatially aggregated data, more precisely, aggregated about road segments, to discover hotspots \cite{shafabakhsh2017gis, aghajani2017applying}. 
The quantity and density of accidents on each road section were combined to identify places where accidents frequently occur on a road \cite{okabe2012spatial}. 
Identifying road accident hotspots and corresponding factors are crucial to determine effective strategies for reducing traffic accidents.

Studies show that the high risk of traffic accidents is influenced by a range of human and vehicle factors such as distracted driving and excessive speed as well as various environmental elements, including inadequate road infrastructure, road type, insufficient signage, and adverse weather conditions \citep{briz2019identification}.
Furthermore, temporal factors such as the time of day, including nights and weekends, have a crucial role in determining the frequency and severity of accidents \citep{shope2008teen, liu2017exploring}. Most of the causes for traffic collisions can be broadly classified into temporal and spatial components.
A series of studies \citep{prasannakumar2011spatio, zhong2014combined, mahata2019spatio} analyze historic data to explain the effects of risk factors and assess the likelihood of events to categorize each factor affecting traffic accidents in space and time. These factors play a crucial role in statistical analysis and in the construction of predictive models, as noted by \cite{kumar2013poisson}.
The study by \citeauthor{liu2017dynamic} considered several spatiotemporal interacting and triggering elements that influence the likelihood of traffic accidents in a particular environment. Their model integrates the traffic shockwave model and GIS road network to analyze the spread of the influence of these incidents.
Depending on the specific interest of such studies, a multi-disciplinary research approach is essential to explore the spatiotemporal dynamics of road traffic accidents with covariates.
When applying spatiotemporal modeling to traffic accidents, identifying important factors has gained increasing interest in the domain of road safety management \citep {williamson1995causes, khulbe2019modeling}.

Numerous factors influence the multifaceted random phenomenon of the spatial and temporal distribution of road accidents. 
For these processes, one can employ either of the following two main mathematical modeling options.
A mechanistic model is one type of model that is used to analyze traffic accidents. 
These models aim to systematically analyze the causal factors of accidents by considering the interaction between the road, driver, vehicles, and various dynamic spatial and temporal factors \citep{vallati2016efficient}. 
The second model is a statistical approach that utilizes historical data to identify the underlying process responsible for generating spatiotemporal occurrences using a probabilistic mechanism. 
The statistical models can further be specified based on primary interests of traffic accident modeling into: models that analyze the impact of risk factors on accident probability \citep{potoglou2018factors, rista2018examining}, predictive models that forecast accidents \citep{gianfranco2018accident}, and spatial and spatiotemporal models that directly estimate accident intensity using point processes \citep{hashimoto2016development, shafabakhsh2017gis}.

Various statistical techniques have been used to analyze the spatial variability of traffic accidents to understand where and when these accidents happen in a region. 
Most of these studies use a Poisson regression models, which assume that the number of accidents in a given area or time period is independent and randomly distributed. 
However, traffic accidents are often clustered in space and time, which means that the Poisson regression model may not be appropriate. 
One way to address this problem is to use Poisson model variations \citep{kumar2013poisson, castro2012latent} (for e.g., negative binomial or Poisson-gamma distribution) in a generalized linear model (GLM) to account for the analysis of spatial variability of traffic accidents, or the tendency of points to be clustered together. 
Another option is to use a zero-inflated Poisson regression \citep{roshandeh2016statistical}, which can account for the fact that some areas or time periods have no accidents. 
However, even these models may not be able to fully capture the complexity of traffic accident data. This is because there are often unobserved factors that can affect the intensity of traffic accidents, such as weather conditions, road conditions, and driver behavior etc. 
These factors can lead to spatial and temporal correlation in the data, which can make it difficult to estimate the parameters of the model. 
For example, \citeauthor{wang2019factors} studied the factors influencing traffic accident frequencies on urban roads and found that accidents occurring at different locations are related, supporting the idea of spatial autocorrelation in traffic accident events. 
One alternative to these models is to use a non-homogeneous Poisson process parameterization.
\citeauthor{Karaganis2006Spatial} used a spatial point process model as a non-homogeneous Poisson process for estimating the probability of occurrence of a traffic accident in different road segments of the national highway of Greece. In their study, they treated the road as a line, even though in reality, it should be viewed as a region.
In general, such an approach uses a stochastic intensity function to capture the spatial and temporal effects on the number of accidents. Several studies \citep{kumar2013poisson, juan2012pinpointing, Karaganis2006Spatial} indicate that stochastic spatial processes are highly effective analytical methods for examining the spatial and spatiotemporal patterns of traffic crashes. However, parameterization in such approach may not have closed form solutions and hence studies often use a Bayesian approach to estimate the corresponding parameters.
 
\citeauthor{ramirez2021spatiotemporal} uses a Bayesian approach based on Log-Gaussian Cox process to model intensity and relative risk of traffic accidents in the city of Bogota, Colombia as spatiotemporal point process by considering a regular 32 by 64 grid to calculate the continuous spatial covariates in each cell.
They used Markov chain Monte Carlo (MCMC) for posterior estimation and showed the capability of the Bayesian formulation to identify factors that increase the risk of accidents though convergence challenge of the the MCMC is inevitable for large data sets \citep{rue2009approximate}.
\citeauthor{cantillo2016factors} employed a GIS-empirical Bayesian methodology to model traffic accidents occurring on the metropolitan roads of Colombia and examine the relationship between urban road accidents and variables related to road infrastructure, environment, traffic volumes and traffic control.
\citeauthor{Karaganis2006Spatial}  also used Bayesian methodology for spatial point processes to investigate the probability of occurrence of a traffic accident in a segment of the national highway of Greek without considering factors for the occurrence of traffic accidents. 
In statistical analysis and prediction of traffic accidents, considering the Bayesian framework is common recently. For example, \cite{chaudhuri2023spatiotemporal,chaudhuri2023spatio} employed a spatiotemporal model to provide predictions of the number of traffic collisions on any given road segment to further generate a risk map of the entire road network.
Their modeling procedure utilized a Bayesian methodology that incorporated Integrated Nested Laplace Approximations (INLA) with Stochastic Partial Differential Equations (SPDE). They applied the methodology to traffic accident data for specific road segments of Barcelona, Spain, during the period 2010--2019 and to traffic accident data of London, UK, during the period 2013-2017.
They improved the SPDE triangulation approach, especially for linear networks by developing a generalized methodology to model spatial and spatiotemporal events in complex spatial structures and compared the results with the barrier model of \cite{bakka2019non}.
However, their study is more applicable only for the spatiotemporal dynamics of geostatistical data and/or for marked point pattern data on a road network. 
\citeauthor{Flaag2023inlalgcp} also considered the same point process model, but only for spatial point pattern data, and tested the model on the locations of 3605 Beilschmiedia pendula Lauraceae trees in a 1000m by 500m plot from a tropical rainforest on Barro Colorado Island, Panama. 
Thus, this study is motivated by the need to consider traffic accident data points as point pattern data, investigate the spatiotemporal dynamics of the process not only for the entire spatial region of Addis Ababa of Ethiopia but also to its road networks separately.
We plan to formulate the Bayesial model with covariate and without covariate for both the entire spatial and road network consideration cases and compare the effects of different covariates to the model.
The use of spatiotemporal modeling techniques can help account for the complex interactions between different factors and their impact on local traffic accidents over time and space. This can also help researchers identify trends and patterns that may not be apparent through simple statistical methods such as kernel density estimation \citep{hashimoto2016development, chaudhuri2023trend}.

%\textcolor{red}{Also, there are new studies on spatiotemporal point processes over networks \citep{moradi2020first}, and \cite{chaudhuri2023trend}. These studies identify patterns in which points influence and connect in space. By collecting information on how often accidents happen and how busy the roads are overall; \cite{chaudhuri2023spatiotemporal} makes it possible to use binomial regression models along with a Bayesian framework to predict accidents on specific road segments. Recently \cite{chaudhuri2023enhanced} discussed the challenges of using INLA and traditional SPDE method in implementing Bayesian spatiotemporal modeling in complex linear networks. } 

This study proposes and evaluates a class of spatiotemporal point process models called Log-Gaussian Cox processes (LGCP) \citep{moller1998log} together with and without covariates for analyzing point patterns in urban traffic accident events. 
As noted by \citeauthor{ramirez2021spatiotemporal},
the LGCP differs from other point process models in that it does not rely on a random contagious process for creating clusters and spatial correlations. Instead, it uses a stochastic intensity function to model the clustering of events, which is influenced by environmental factors and other conditions.
In this study, we model the log-intensity of traffic accidents with and without fixed and random effect components that incorporate possible covariates and spatial correlation information respectively.
We model covariates such as population density and distance of accident from crowds (like nearest schools, markets, bus stations, worship places) as fixed effects and the spatial correlation of events as random effects through Mat\'ern field.
In general, we use a Bayesian approach by considering the Poisson model for the likelihood of predicting the underlying process using the spatiotemoral intensity estimates on the entire spatial domain and on the road networks separately, which permits to identify critical accident zones both spatially in the entire region and on the road network and how these locations change with time. 
We implement the proposed model to Addis Ababa, Ethiopia traffic accident data during the period 2016--2019. 

The remainder of the document is organized as follows: In Section \ref{sec_data_settiongs}, we describe the data sources, data integration, georeferntiation and techniques used to combine them into a single dataset. In Section \ref{sec1}, we present the statistical methodology for analyzing data and estimating the spatiotemporal point process model. In Section \ref{Sec_results}, we present  and analyze the result of model implementation by estimating the spatiotemporal traffic accident intensity map, and Section \ref{Sec_conclusions} concludes the study with a discussion.

\section{Data settings}\label{sec_data_settiongs}
\begin{figure}[H]
	\centering
	\includegraphics[width=1\columnwidth]{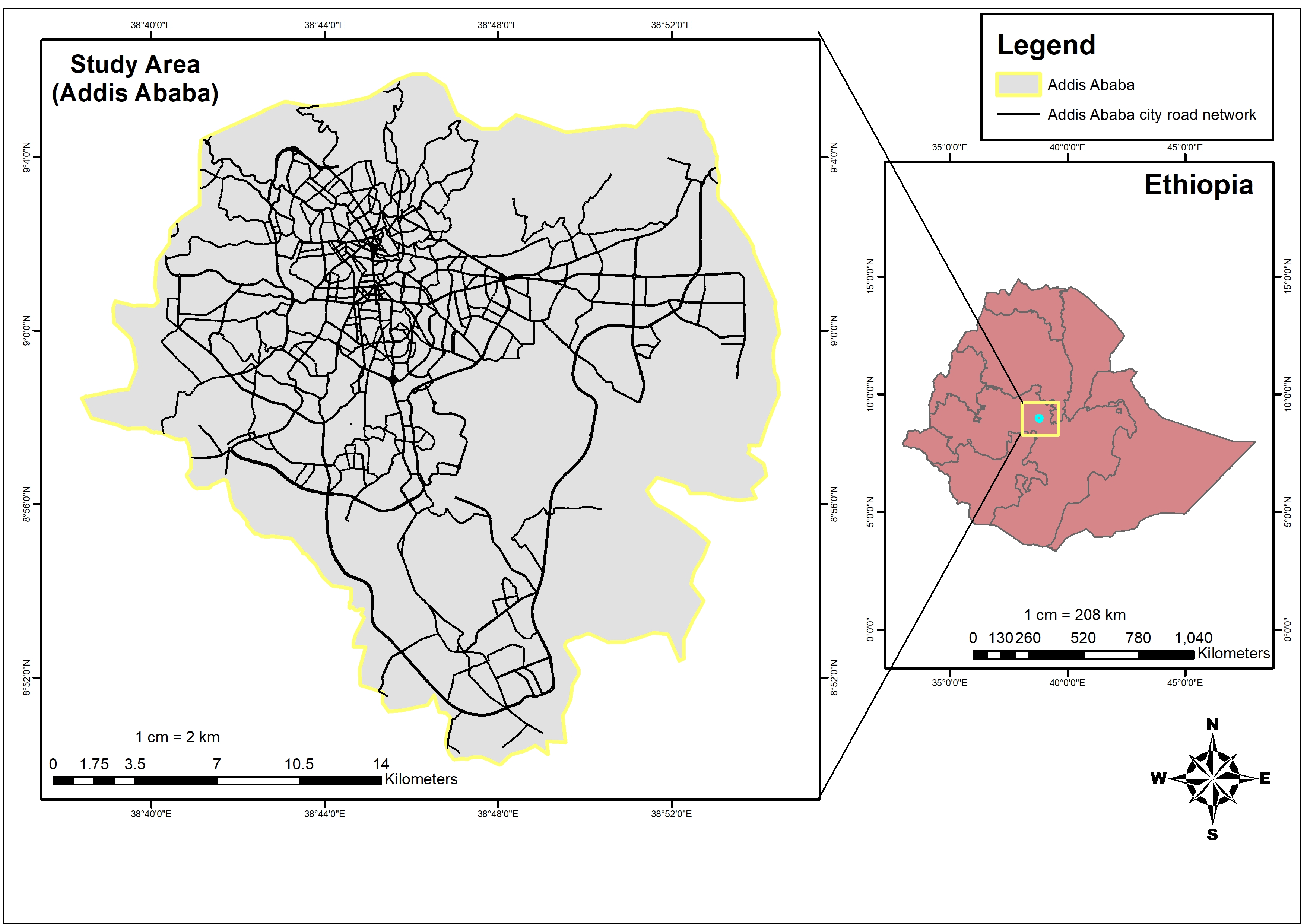}
	\caption{The polygon (left) inside the box with yellow boundary is the geography of the City Addis Ababa (Ethiopia), while the black segments represent the selected of specific category of street road network used in this study, whereas the whole map is used to show and locate the study-area with respect to Ethiopia.}
	\label{sa}
\end{figure}

Addis Ababa is the largest metropolitan area and capital city of Ethiopia where traffic accident is a critical issue. The city has an area of 528 square kilometers with 5.353 million residents, or roughly 10138 people per square kilometer, according to Ethiopian Statistical Service \citep{csaeth}.
%\footnote{\url{http://www.csa.gov.et/}}. 
Figure \ref{sa} shows the location of Addis Ababa. The boundary of the municipality is highlighted in yellow. The city is also a significant hub for Ethiopia's finances, economy, culture, and serving as the nation's commerce hub. 
In this study, we consider the entire area of Addis Ababa with selected road types. The road network was made up of over 12,000 road segments (edges) with a total length of 1218.172 km, with available covariates for analysis. Due to its significance as the most populous city in Ethiopia and the location of over 60\% of all car crashes, Addis Ababa was chosen as the study site. In this study, we used a 4-years traffic fatality data obtained from Addis Ababa traffic management office
%\footnote{\url{https://www.tma.gov.et/traffic-management-strategy-for-safe-and-acceptable-traffic-movement-report/}} 
during the period 2016 -- 2019 \citep{aatm}.  
These data attributes include accident ID, location, type of accident, date of year, and driver information.
Since the data was recorded with location of specific places or buildings etc for each event, we used Awesome Table feature from the Google Sheets \citep{googleawesome}
%\footnote{\url{https://support.awesome-table.com/hc/en-us/sections/360000012309-Geocode}}, 
to associate a corresponding coordinate reference system for each spatial locations recorded.

\begin{figure}[H]
	\centering
	\includegraphics[width= 1\textwidth]{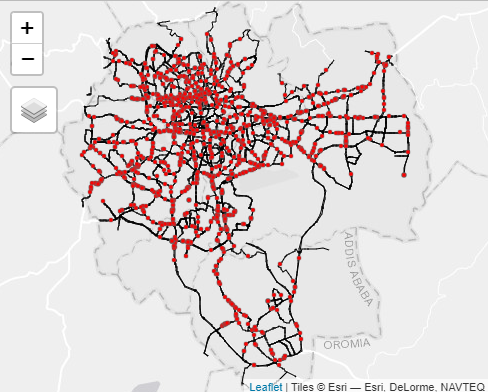}
	\caption{Selected road network of Addis Ababa used for study together with 1474 traffic death locations from the year 2016 to 2019}
	\label{d2}
\end{figure}

We projected all crash locations onto the nearest road segment of the road network, used these locations as response variables for the statistical model, and calculated the shortest distances from the nearest bus stop, market place or street market, schools, hospitals, restaurants, and places of worship for each accident location. We use these distances and population density obtained and extracted from WorldPop
%\footnote{\url{https://www.worldpop.org/datacatalog/}} 
as spatial covariates in our datasets \citep{worldpop2}. 
Figure \eqref{d2} highlights the road network, along with death locations in red.
The R package osmdata \citep{padgham2017osmdata} and OSMnx \citep{boeing2017osmnx} of Python are used to retrieve the road network from the open street map (OSM) servers
%\footnote{\url{https://www.openstreetmap.org/copyright}} 
repository \citep{OpenStreetMap}. We used the primary tag highway (applied for any type of street) to retrieve the entire OSM street network for the study area, and then filter the suitable highway types of primary, secondary, tertiary and trunk roads for our study. 

\section{Statistical modeling framework}\label{sec1}
Consider spatial /or spatiotemporal point process $\mathcal{X}$ and its partial realization as $\mathbf{x}$ over a region $\mathcal{W}$, where $\mathcal{W}$ denote either a spatial region $S\subset\mathbb{R}^d$ or a spatiotemporal region $S\times T\subset\mathbb{R}^d\times\mathbb{R}$. 
Then, the realizations of the point process $\mathcal{X}$ is a collection of points $\{s_i\}_{i=1,\cdots, n}$ or $\{(s_i,t_i)\}_{i=1,\cdots, n}\subset \mathcal{W}$.
A realized  $\mathbf{x}$ is also called a point pattern and its elements are the events \cite{daley2003introduction,gelfand2010handbook,diggle2013statistical}. 
Such point patterns are modeled with Poisson processes that are characterised by their corresponding intensity functions denoted by $\lambda(\cdot)$ and defined as the average number of events occurred per unit area and/or time depending on the point of interest.
Hence, the main characteristics of the spatiotemporal distribution of point patterns are explainable by the intensity function of the Poisson process, which governs the distribution of the point process $\mathcal{X}$.

A Poisson process with intensity function $\lambda(\cdot)$ satisfies the following two properties \citep{moller2003statistical}
\begin{enumerate}
	\item For any (Borel) subset $A\subset S\times T$, the number of events in $A$ denoted by $N(A)$ follows a Poisson distribution with mean $\int_A\lambda(s,t)dsdt$. In this way, we are assuming that the number of observations that occur in disjoint sets are independent.
	\item Given the number of events in $A$, i.e., $N(A) =n$, those events are independent and identically distributed with probability density proportional to $\lambda(s,t)$. %(i.e., $$\frac{\lambda(s,t)}{\int_A\lambda(s,t)}$$)
\end{enumerate}
The distribution function for the Poisson point process for a bounded region $\mathcal{W}\subset\mathbb{R}^d\times\mathbb{R}$ is given by

\begin{equation}\label{01}
	\pi(\mathbf{x}|\lambda(s,t))\propto \exp\bigg(-\int_{\mathcal{W}}\lambda(s,t)dsdt \bigg)\prod_{(s_i,t_i)\in\mathbf{x}}\lambda(s_i,t_i)
\end{equation}

where $(s_i, t_i)$, are respectively the locations in space-time of the observed events as given by \cite{cressie2015statistics}. 
This can be expanded by introducing parameters that are used to define the intensity, most typically using a log-linear model with covariates and/or spatial correlation information that can be expressed by a polynomial or spline function of the geographical coordinates. 
In a Poisson process, the intensity function is usually assumed to be fixed parameter. 
This is frequently oversimplified in applications of spatial or spatiotemporal point process models since geographic heterogeneity can be generated by other random processes with their own characteristics from which inferences are required. Thus, the intensity function shall best be modeled as a random field $\lambda(s,t)$ \citep{moller2003statistical}. 
This creates a flexible set of point processes known as Cox or doubly stochastic Poisson processes. 
Typically, these processes simulate aggregation in point patterns caused by visible or undetected environmental change.
A Cox process must satisfy the two Poisson process criteria, where the intensity function might be random. 
Cox processes are often used to model event clustering.

Here, we model the Spatiotemporal point process as log-Gaussian Cox processes (LGCP) \citep{moller1998log} as
\begin{equation}
	\log \lambda(s,t) = \eta(s,t), 
\end{equation}
where $\eta(s,t)$ is a Gaussian Process.
In this study, we are interested in modeling the log-normal intensity as a linear combination of fixed effect components describing the covariates of the process of interest and a random effect describing the spatial correlation of the occurrence of the events in a region. 
We also plan to consider the model without covariates by simply considering the later component to see effect of covariates in the model.

The decomposition proposed in this study is particularly valuable for examining potential variations in the temporal and spatial patterns of occurrences of events with spatiotemporal factors within a Bayesian framework consideration. 
The method captures fluctuations in the temporal correlation and variations in the geographical distribution pattern using the time-varying spatial random effect. 
The Bayesian hierarchical spatiotemporal point process model we use follows the classical approach for LGCP and adopts the notation from \cite{simpson2016grid} and \cite{Flaag2023inlalgcp}. In this model, the logarithm of the intensity $\lambda(s,t)$ is modeled as a Gaussian process. The model includes linear spatial covariates $Z(s, t)$ and a zero mean additive Gaussian spatiotemporal random field $\xi(s, t)$ and is given as follows
\begin{equation}\label{03}
    \begin{split}
	\mathbf{x} \mid \lambda(s,t) &\propto \text{Poisson}\big(\lambda (s,t)\big)\\
	\log\lambda(s,t)) &= \eta(s,t)\\ 
	\eta(s,t) &= \beta_0 + Z'(s,t)\beta + \xi(s,t),
	%\xi(s,t) &= \phi \xi(s,t-1) + \omega(s,t)
    \end{split}
\end{equation}
where $\mathbf{x}$ is the observed points, $\beta_0$ is the intercept, and $Z(s,t)$ is the associated purely spatial covariate at location $s$ which may not vary with time $t$. 

For practical implementation reasons, spatial covariates are projected onto the same computational function space as the latent field. The $\xi(s, t)$ is the spatial random effects represented by the Gaussian process $\omega(s, t)$ continuously projected in space independent in time given by
\begin{equation}\label{two}
	\xi(s,t) = \phi \xi(s,t-1) + \omega(s,t)
\end{equation}
for $t = 2,\cdots,T, \phi$ is is the temporal correlation parameter, where $|\phi| < 1$ and $\omega(s,1)$ derives from the stationary distribution $\mathcal{N}\big(0,\frac{\sigma^2_{\omega}}{1 - \phi^2}\big)$. Additionally, $\xi(s,t)$ is assumed to be zero-mean spatiotemporal Gaussian random fields model in space that follows an AR(1) process over time with dependent parameter $\phi$ \citep{lindgren2011explicit, cameletti2013spatio} that refers to the latent spatiotemporal process (i.e., the true unobserved level of traffic accident death at location $s$ in time $t$). In this case, we assume the spatiotemporal Gaussian random field are independent at each time step $t$; and that changes in time with AR(1) constructions in discrete, and equally spaced time. Moreover, $\omega(s,t)$ is a zero-mean Gaussian random field, is assumed to be temporally independent and is characterized by a zero mean and a spatiotemporal covariance function given by
\begin{equation}\label{three}
	Cov\big(\omega(s_i,t),\omega(s_j,t')\big) = \begin{cases}
		\sigma^2_{\omega}\mathcal{C}(r)  &  \text{ if $ t = t'$}\\
		0 &  \text{if $t \neq t'$} 
	\end{cases} ~~~~~\text{for $s_i\neq s_j$.}
\end{equation}
The purely spatial correlation function $\mathcal{C}(r)$ depends on Euclidean distance $r = ||s_i-s_j||$ between locations $s_i$ and $s_j$ and it is defined by a  Mat\'ern correlation function that can be given as
\begin{equation}\label{four}
	\mathcal{C}(r)=\frac{1}{2^{\nu-1}\Gamma(\nu)}(\kappa r)^\nu K_\nu(\kappa r)	
\end{equation}
where $K_\nu(\cdot)$ is the modified Bessel function of second kind and order $\nu >0$, and $\nu$ is a smoothing parameter which measures the degree of smoothness of the process \& its integer value determines the mean square differentiablity of the process \cite{lindgren2011explicit}. Furthermore, the parameter $\kappa$ is a positive scaling factor that establishes a relationship between the range $\rho$. 

When considering covariance structures, the Mat\'ern covariance function is particularly useful for modeling spatial data. We might acquire specific instances of that function, such as the Gaussian and exponential covariance functions, to demonstrate its adaptability. In addition, it provides accurate approximation properties for other spatial dependence models, as stated in \cite{Illian2012toolbox}. The marginal variance $\sigma^2_{\omega}$ is given by 
\begin{equation}\label{02}
	\sigma^2_{\omega} = \Gamma(\nu)\Gamma\left(\nu + \frac{d}{2}\right)^{-1}\tau^2 (4\pi)^{-\frac{d}{2}}\kappa^{-2\nu},
\end{equation}
where $\tau$ is a scaling parameter and $d$ (in our case $d = 2$) is the spatial dimension. As stated in \citeauthor{lindgren2011explicit}, we choose to parameterize the covariance function using $\log \tau$ and $\log \kappa$ by
\begin{align}
	\log\tau &= \frac{1}{2}\left(\frac{\Gamma(\nu)}{\Gamma(\alpha)(4\pi)^{0.5d}}\right)-\log\sigma_\omega -\nu\log\rho\nonumber\\
	\log\kappa &=\frac{1}{2}\log 8\nu - \log\rho,
\end{align}
where $\rho = \frac{\sqrt{8\nu}}{\kappa}$. Depending on the value of $\nu$, this approach has the benefit of only requiring the estimation of two parameters.

Given a bounded region $\mathcal{W}\subset\mathbb{R}^2$ and $\mathbf{x} =\{\mathbf{x}_1,\cdots, \mathbf{x}_T\}$ be the observed points from (a conditionally) Poisson process with intensity $\lambda(s,t)$ at time $t=1,\cdots,T$, where $\mathbf{x}_t = \{(s_i,t)\}$ for  $t = 1,\cdots, T, \text{and}~ i=1,2,\cdots, n_t$ where $n_t$ is the observed points at time $t$. Then, the Poisson process likelihood \citep{kingman1992poisson, moller1998log} given in Equation \eqref{01} is reduced into
\begin{equation}\label{04}
	L(\mathbf{x}|\lambda)\propto \exp\left(-\int_\mathcal{W}\lambda(s,t)ds\right)\prod_{t=1}^{T}\prod_{i=1}^{n_t}\lambda(s_i,t).
\end{equation}

Due to the doubly stochastic property of the intensity function, the likelihood in Equation \eqref{04} is analytically intractable because of the stochastic integral. The most known technique to Bayesian fitting of LGCP models but is also common in frequentist analyses, is to grid the domain and model the induced Poisson counts. This still creates a problem because the number of grids determines the size of the covariance matrix whose inverse and determinant will be required in Bayesian computations \citep{moller1998log}. A computationally efficient approximation to the Poisson process likelihood that requires the intensity function only to be evaluated at the locations of observed events and at mesh nodes was introduced by \cite{simpson2016grid}. Thus, the SPDE approach can be employed to model the intensity function and the same nodes reused in the evaluation of the Poisson process likelihood. 

Since the LGCP includes a Gaussian random field (GRF), efficient computation for GRF is crucial when working with LGCP models. This GRF imposes a dense covariance matrix on the latent variables \cite{blangiardo2015spatial}. Suppose $\omega(s)$ be a Mat\'ern field of second order isotropic and stationary GRF with Mat\'ern covariance function given in Equation \eqref{four} and suppose we have a realization of $\omega(s)$ at $n$ spatial locations $s_1,\dots, s_m$. The primary benefits of the SPDE approach are to find a Gaussian markov random field (GMRF) \citep{rue2005gaussian}, with a sparse precision matrix using a neighborhood structure, that efficiently describes the Mat\'ern field which is best computational expense when making inferences. Accordingly, it is conceivable to make inference \citep{lindgren2011explicit}. As shown by \citeauthor{whittle1963stochastic}, a GRF with  Mat\'ern covariance matrix can be represented as a solution of the continuous domain SPDE of the form
\begin{equation}\label{11}
	(\kappa^2 - \Delta)^{\frac{\alpha}{2}}(\tau \omega(s)) = W(s),~ s\in\mathbb{R}^d,~\kappa>0,~\alpha = (\nu + \frac{d}{2}),~\nu>0.
\end{equation}
Here, $\Delta = \sum_{i=1}^{d}\frac{\partial^2}{\partial s_i^2}$ is Laplace operator, and $W(s)$ is a spatial white noise with variance 1. The stationary solution $\omega(s)$ is a Gaussian field having a Mat\'ern covariance function with precision (inverse variance) $\tau$, scaling parameter $\kappa$ (inversely proportional to the effective spatial range $\rho$), and smoothness parameter $\nu$. The smoothness parameter $\nu$ considered in the Mat\'ern covariance function corresponds to integer values of $\alpha$ \citep{lindgren2011explicit}. 

The finite element method (FEM) can be used to find the approximated solution to the SPDE given in Equation \eqref{spde} by dividing the entire spatial domain into a set of non-intersecting triangles, creating a triangulated mesh with nodes and piecewise linear basis functions defined on each triangle. Then, the continuously indexed Gaussian field $\omega(\cdot)$ is represented as discretely indexed GMRF as 
\begin{equation}\label{spde}
	\omega(s,t)\approx  \tilde{\omega}(s,t) = \sum_{j=1}^m w_j\psi_j(s,t),
\end{equation}
where $m$ is number of vertices of the triangulation, $\{\psi_j(s)\}_{j=1}^m$ is piecewise linear basis functions defined for each node on the mesh, and $\{w_j\}_{j=1}^m$ are zero mean Gaussian distributed weights. The spatial field value at each vertex of the triangle is defined by the weight $w_j$. The values inside the triangle, which are used to create a barycentric vertex, are calculated by linear interpolation using the triangle's vertices \citep{Flaag2023inlalgcp}. The way the weights are spread out ($\{w_j\}$) is the same as the answer to the SPDE in Equation \eqref{11} for a set of test functions ($\{\psi_j\}$). The equation \eqref{spde} presents a finite element form of the SPDE technique. Using the Markovian property, this equation establishes a connection between the Green's function $\omega(s,t)$ and the GMRF defined by the Gaussian weights ${w_i}$. The precision matrix of the GMRF $w = (w_1, \cdots, w_n)'$ is determined by $\kappa^2$ \citep{lindgren2011explicit}. Here, $\alpha$ and $\nu$ are integers satisfying $\alpha \geq 1$ and $\nu \geq 0$, with $\alpha = \nu+1$. The multivariate Gaussian vector consists of the latent field values at the nodes. This provides a clear correspondence between the parameters ($\nu, \kappa$) of the covariance function for Gaussian Random Field (GRF) and the elements of the precision matrix $Q$ of the Gaussian Markov Random Field (GMRF) $w$, with a computational cost of $\mathcal{O}(n^{\frac{3}{2}})$ for any triangulation \citep{cameletti2013spatio}. By substituting the GRF $\omega(s, t)$ with the GMRF approximation $\tilde{\omega}(s, t)$ in Equation \eqref{03} and by approximating the integral in the likelihood of Equation \eqref{04} using a quadrature rule \cite{simpson2016grid}, the resulting approximate likelihood consists of $T\times(m + n_t)$ independent Poisson random variables. Here, $m$ represents the number of mesh vertices, and $n_t$ is the number of observed points at time $t$. 
For the model without covariates, we consider the log-intensity given in Equation \eqref{03} as the intercept plus the SPDE term only without the covariates term $Z'(s,t)\beta$.

\subsection{Bayesian inference}
\citeauthor{simpson2016grid} showed the LGCP formulation within the Bayesian hierarchical modelling framework. Thus, the estimates of the corresponding posteriors can best be approximated through the INLA approach of \citeauthor{rue2009approximate} to avoid the convergence challenges of MCMC methods for the class of latent Gaussian models \cite{rue2009approximate, lindgren2015bayesian}. 
Therefore, before considering a posterior distribution of the parameters $\boldsymbol{\beta}, \boldsymbol{\theta}$ and the process $\boldsymbol{\xi}$, the likelihood of the observed point patterns $\mathbf{x} $ with intensity $\lambda$, described by Equation \eqref{03}, is defined through the conditional density as
\begin{equation}
	\pi(\mathbf{x} \mid \boldsymbol{\beta}, \boldsymbol{\theta}, \boldsymbol{\xi}) = \prod \pi(\mathbf{x}_i \mid\boldsymbol{\beta}, \boldsymbol{\xi})
\end{equation}
where $\boldsymbol{\theta}$ are parameters for the latent spatiotemporal process $\boldsymbol{\xi}$ and the equality holds due to the conditional independence assumption between $\mathbf{x}$ and $\boldsymbol{\theta}$ given $\boldsymbol{\xi}$. This likelihood is equivalent to Equation \eqref{04}. 
Thus, using Baye's theorem the conditional posterior probability of $(\boldsymbol{\beta, \theta, \xi})$ conditioned on the point patterns is proportional to the product of the likelihood and the prior distribution and is given as 
\begin{equation}
	\pi(\boldsymbol{\beta, \theta, \xi} \mid \mathbf{x}) \propto \pi(\mathbf{x} \mid \boldsymbol{\beta}, \boldsymbol{\xi}) \pi(\boldsymbol{\beta, \theta, \xi})
\end{equation}
where, $\pi(\boldsymbol{\beta, \theta, \xi})$ accounts for the prior distribution. We use the INLA for the estimation of the conditional posteriors of each parameters.
The INLA approach makes prior sensitivity analysis and model comparisons more viable by avoiding sampling, unlike MCMC, allows for rapid model fitting even in the presence of massive datasets \cite{rue2009approximate}. 

\begin{figure}[H]
	\centering
	\includegraphics[width=.9\linewidth]{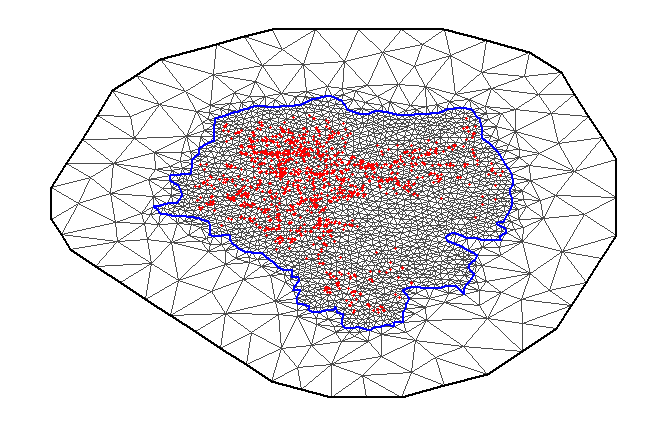}
	\caption{The observed spatial traffic death locations  (red dots) and the corresponding triangulation mesh used for the SPDE model with an inner and outer boundary (blue and black lines). The inner boundary delimits a high-resolution zone covering the study region, while the outer boundary delimits an extension zone with lower resolution to prevent the boundary effects of SPDE conditions in the study region}
	\label{spat.mesh}
\end{figure}

\begin{figure}[H]
	\centering
	\includegraphics[width=1\linewidth]{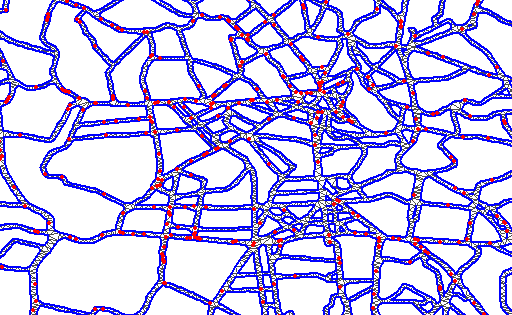}
	\caption{Partial network mesh of selected road from the whole road shown in Figure \ref{d2} in blue and observed data locations highlighted as red points.}
	\label{SPDE_net.mesh}
\end{figure}

The INLA-SPDE needs mesh construction, also known as triangulation, to connect discrete event locations and to figure out a continuous process in space \citep{rue2017bayesian}. 
In this work, we use two study domains for building the mesh; the complete study area, Addis Ababa, which is shown in Figure \ref{spat.mesh} and only the road networks of Addis Ababa only as shown in Figure \ref{SPDE_net.mesh}. 
As stated in \citeauthor{okabe2012spatial}, traffic accident are limited to specific locations on the road networks, and the model that uses the whole spatial region forecast of results in places without a road network, where traffic accidents are unlikely to occur, may seam not practical, considering the road networks only sounds reasonable. 
However, some of the covariates considered might be not covariates on the road directly, like the population density, instead in the entire region and it might be reasonable to consider both the complete spataial domain and the road network. 
Hence, in this study, we prefer to consider both mesh construction and implementation of the model for the entire region and for the road networks separately (of Addis Ababa) and compare the model outputs, and effects of the covariates on the two considerations.
We follow \citeauthor{chaudhuri2023spatio} approach to construct a mesh on the road network. 
The approach gives rise to the rationale behind reasonably accurate formulation of SPDE triangulation on the road networks only. 
To create buffers for particular road segments, we utilized the R package rgeos \citep{bivand2017rgeos} to construct a buffer region for the selected road network with 30 meters added buffer. 
The SPDE network mesh, together with the traffic accident locations, is depicted in Figure \ref{SPDE_net.mesh}, which has 16325 mesh vertices. 

We build an SPDE model, using \textit{inla.spde2.pcmatern} function from R-INLA package, and specify penalized complexity (PC) priors \citep{simpson2017penalising} for parameters of the Mat\'{e}rn field. 
Following the approach by \cite{bakka2018spatial}, we define the PC prior considering the range values of the domain of event locations in the study area to have a median around half of the study area.
These priors are strong since they don't affect the outcomes, and the scale also establishes the size of the effects and makes it easier to understand the results \citep{simpson2017penalising}.

We started by modeling intercept and SPDE (spatial random effect) on both domains to see the intensity estimate of the model without covariates as a baseline model.
This is beneficial to identify high intensity places, which have high rates of traffic accidents. 
We then, incorporate the possible covariates such as the population density obtained from WorldPop (that is, the density is defined as approximate population per the area of the study region), and the distances from schools, bus stations, market places, hospitals, restaurants and worship places inside the study region as stated in Section \ref{sec_data_settiongs}. 
The study is limited to these covariates only due to unavailability of other covariates data like the road types and their characteristics, and associated climatic information.
All analyses are carried out using the free R software (version 4.4.1). The results for the intensity map and those parameters from the fixed and random effects are given in the result section.

\section{Analysis and Results}\label{Sec_results}
In this section, we present the findings of the methodological strategy discussed in the previous sections and discuss the model fitting, validation, and predicted intensity maps spatially and on the road network. 

\subsection{Model Fitting}
The proposed model with covariate and without covariate has been fitted to the traffic accident datasets of Addis Ababa, Ethiopia, for the years 2016 -- 2019. 
To fit all the models discussed in Section \ref{sec1}, we used the R-INLA package \cite{rue2009approximate}.
We implemented the methodology, considering the spatial mesh for the entire study region (as shown in Figure \ref{spat.mesh}) and the road network mesh (as its parts is shown in Figure \ref{SPDE_net.mesh}) separately, as it was done by \citeauthor{chaudhuri2023spatio} for marked point pattern data, but for point pattern data in our case. 

We have implemented the model with covariate and without covarite for both mesh categories and covariate combinations. 
We have assessed the performance of each model using the Watanabe-Akaike information criterion (WAIC) and the deviance information criterion (DIC), which are known to balance model complexity and accuracy \citep{spiegelhalter2003bayesian}. 
Despite being more complicated, according to \citeauthor{blangiardo2015spatial}, models with a smaller DIC and/or WAIC fit the sampled data better. The study used the Poisson point process model with and without covariates to model the spatiotemporal structure of traffic accidents for the entire spatial region and for the road networks consideration cases of Addis Ababa . 
The goodness-of-fit summary findings (DIC, WAIC) for the best-fit models with covariates and without covariates  for both the entire spatial region and for the road networks SPDE network mesh consideration are given in Table \ref{DIC}. 
The findings showed that including covariates in the model improves its performance for both the entire spatial region and the road networks SPDE mesh considerations. The model achieved lower values of DIC and WAIC when covariates were taken into account. 

Following the performance of the model with covariates, in the subsequent Sections \ref{subsec:entire_spatial_results} and \ref{subsec:roads_spatial_results}, we have given the corresponding model results and discussions. 
We have given the intensity estimates obtained from the model without covariates in Figures \ref{fig:model-spatial} and \ref{fig:model-road} of the Appendix I for illustration purposes. Both model outputs are found to be reasonably good in terms of determining the accident hotspots though the model with covariate case outputs showed better goodness of fit as the DIC and WAIC indicated. 

\begin{table}[H]
	\caption{DIC, WAIC values for the fitted models}  
	\label{DIC}
	%\small
	\centering
	\begin{tabular}{lccccclcccccccr}
		\toprule\toprule
		\textbf{Model} &&&&&& \textbf{Mesh used} &&&& \textbf{DIC} &&&& \textbf{WAIC} \\ 
		\midrule[2pt]
		\multirow{2}{*}{Without covariates} &&&&&& Spatial region &&&& 1142 &&&& 1311 \\
		&&&&&& Road network &&&& 11253 &&&& 10251\\
		\midrule
		\multirow{2}{*}{With covariates} &&&&&& Spatial region &&&& 1095 &&&& 1258 \\
		&&&&&& Road network &&&& 10107 &&&& 9952\\
		\bottomrule[2pt]
	\end{tabular}	
\end{table}

\subsection{Model outputs for the entire spatial domain}\label{subsec:entire_spatial_results}
\begin{figure}[H]
	\centering
	\includegraphics[width = 1\linewidth]{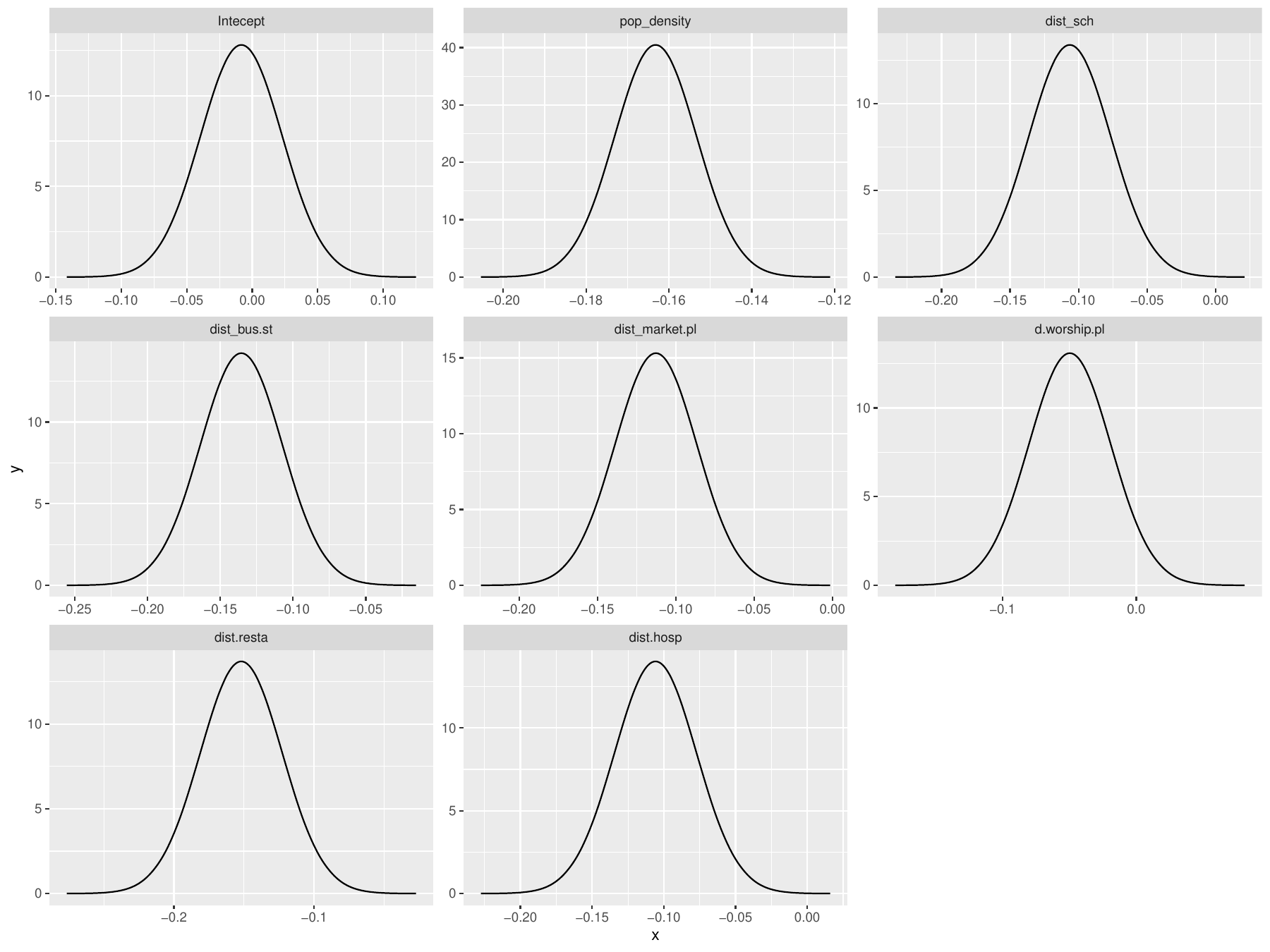} 	\caption{Marginal posterior distributions of fixed effect coefficient from the spatial model with covariates.}
	\label{para}
\end{figure}

\begin{figure}[H]
	\centering
	\includegraphics[width=1\linewidth]{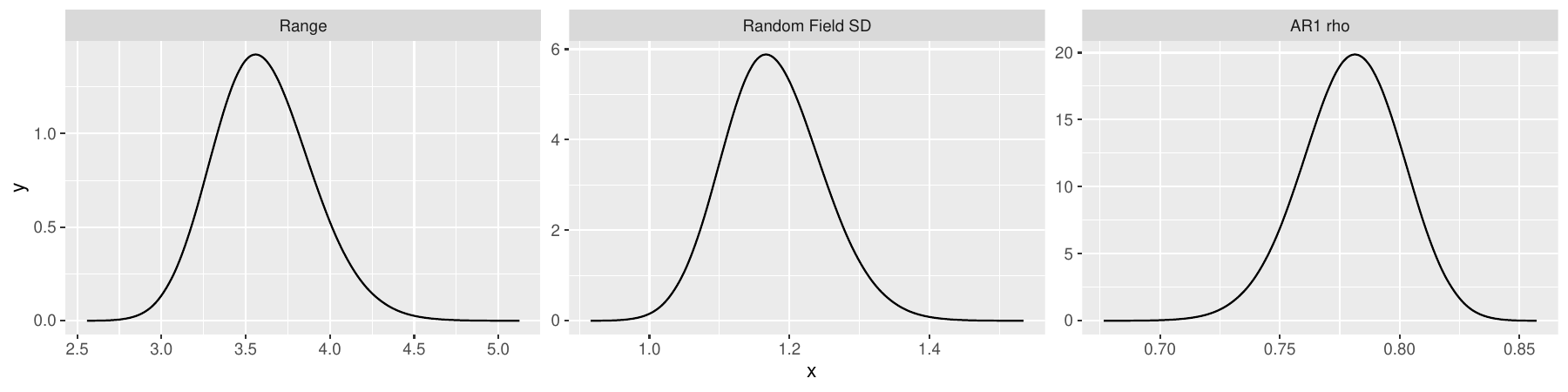}  
	\caption{Marginal posterior distributions of hyperparameters for the spatiotemporal random field of the model using spatial regional mesh.}
	\label{hyper.para.cov}
\end{figure}

The marginal posterior distribution of the parameters for the fixed and random effects included in the model for the entire spatial domain consideration case are depicted in Figures \ref{para} and \ref{hyper.para.cov} respectively. 
Particularly, in Figures \ref{para} we showed the marginal posterior distributions of all fixed effects such as the intercept, the population density, distance from school, from nearest bus station, from supermarkets, from restaurants and from worship places for the entire spatial region consideration. 
Additionally, Table \ref{mean.par} showed the expected values of these fixed effect parameters with corresponding 95\% of credibility for the entire spatial region consideration case.
The results indicated that most of the covariates has negative effect in the model and hence they have inverse relation with the number of traffic accidents and vice-versa, except for the effect of the intercept. 
That is, for example, when the distance from each covariate is small, the number of traffic causality is relatively high in our case when we consider the entire spatial region.
The effect of the intercept to the traffic accident is seen negligible for the entire spatial region consideration case.  
In particular, the results in Table \ref{mean.par} revealed that the covariate associated to the population density, the distance from the nearest restaurant and the distance from the nearest bus station has the highest negative mean values in order, which indicated stronger negative influence on the model relative to the other covariates.

\begin{table}[H] 
	\caption{Credible intervals and marginal posterior mean of parameters of fixed effects for the spatial models with covariates.}
	\label{mean.par}  
	%\small  
	\centering 
	\begin{tabular}{lccccccccccr}  
		\toprule\toprule
		\textbf{Regression parameters for covariates} &&&&&& \textbf{Mean} &&&&& \textbf{Credible intervals}\\
		\midrule[2pt] 
        Intercept ($\beta_0$) &&&&&& -0.007 &&&&& (-0.069, 0.054) \\
		Population density ($\beta_1$) &&&&&& -0.163 &&&&& (-0.183, -0.144) \\
		Distance from nearest school ($\beta_2$) &&&&&& -0.106 &&&&& (-0.165, -0.048)  \\
		Distance from nearest bus station ($\beta_3$) &&&&&& -0.136 &&&&& (-0.191, -0.080) \\
		Distance from nearest market places ($\beta_4$) &&&&&& -0.113 &&&&& (-0.191, -0.061) \\
		Distance from nearest worship places ($\beta_5$) &&&&&& -0.050 &&&&& (-0.110, 0.010) \\ 
		Distance from nearest restaurant ($\beta_6$) &&&&&& -0.152 &&&&& (-0.209, -0.095)  \\
		Distance from nearest hospitals ($\beta_7$) &&&&&& -0.106 &&&&& (-0.162, -0.050) \\ 
		\bottomrule[2pt]  
	\end{tabular} 	
\end{table}  

\begin{figure}[H]
	\centering
	\includegraphics[width=1\linewidth]{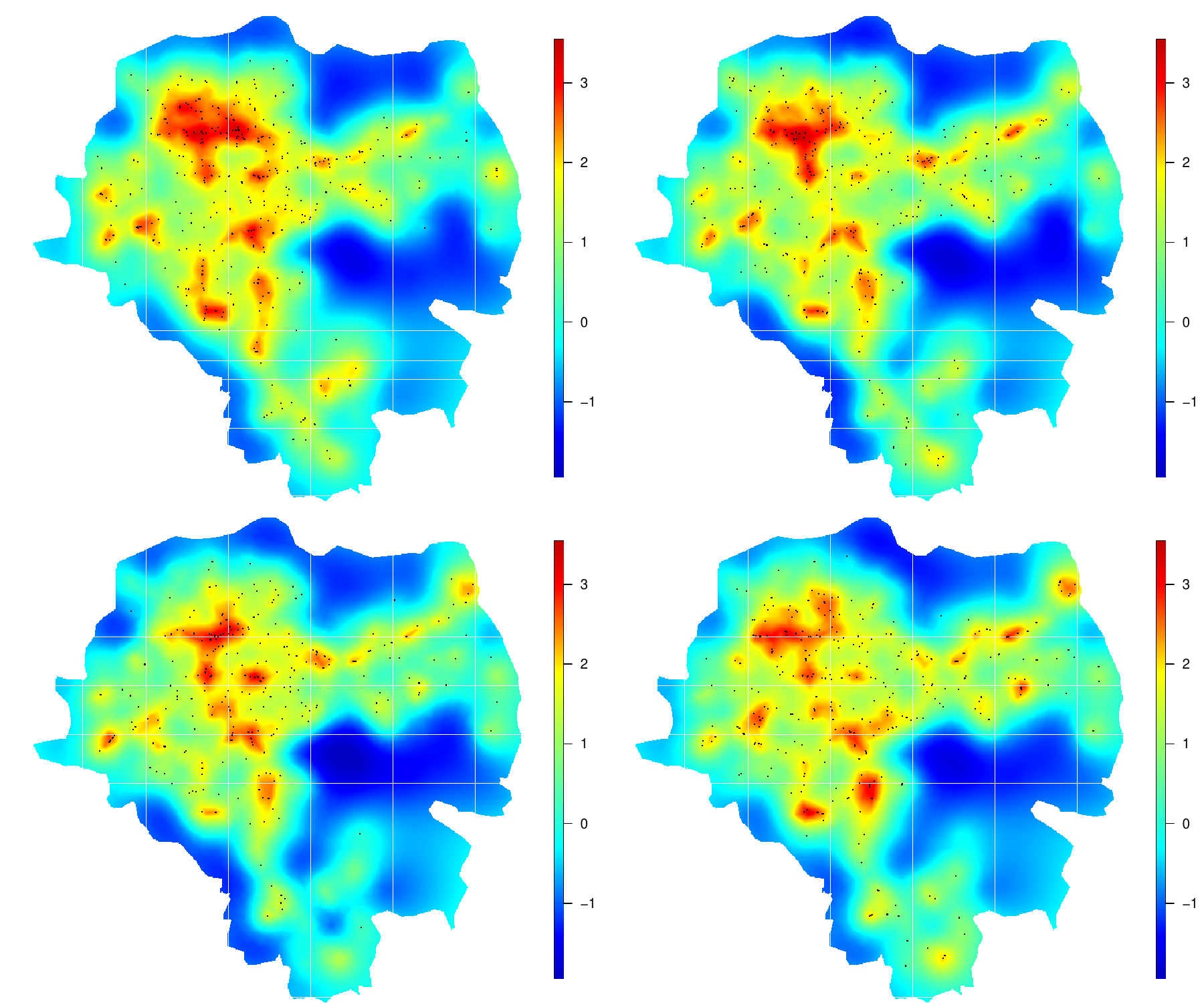}  %spde021.c
	\caption{Maps of the log-intensity function posterior mean for the traffic accident data: years are ordered from the top row to the bottom row, and then from left to right in each row in the sequence 2016 -- 2019 for the model with covariate case.}
	\label{st.BM.int}
\end{figure}

In Figure \ref{st.BM.int}, we have given the log-normal mean of the intensity function estimates from the marginal posterior distribution of the spatiotemporal random effect together with the observation in black points for the entire spatial region consideration case.
The model fit here indicated how well the proposed model is able to describe the traffic accident distribution for each year during the period 2016 -- 2019 in Addis Ababa.
The spatial intensity map here further showed a similar spatial traffic accident trend for each year, with higher traffic accidents in the Northern central part of the city.
We obtained the average temporal correlation of the log-intensity to be 0.78 indicating the existence of a similar traffic fatality trend in space and time from 2016 to 2019.  
The results obtained clearly showed the proposed model's capability to identify traffic accident hotspots, which can guide one to chose safe routes and give insight to decision-makers to prioritize road safety measures and other necessary interventions accordingly.
We have given the marginal posterior distributions of the spatial effect hyperparameters: the range, standard deviation, and temporal correlation in Figure \ref{hyper.para.cov} for illustration purpose.
The average standard deviation of the model output for the random effect is found to be 1.15 indicating the performance of the model for such study in this particular case.

\subsection{Model outputs for the road network of Addis Ababa}\label{subsec:roads_spatial_results}
\begin{figure}[H]
	\centering
	\includegraphics[height = 0.75\linewidth]{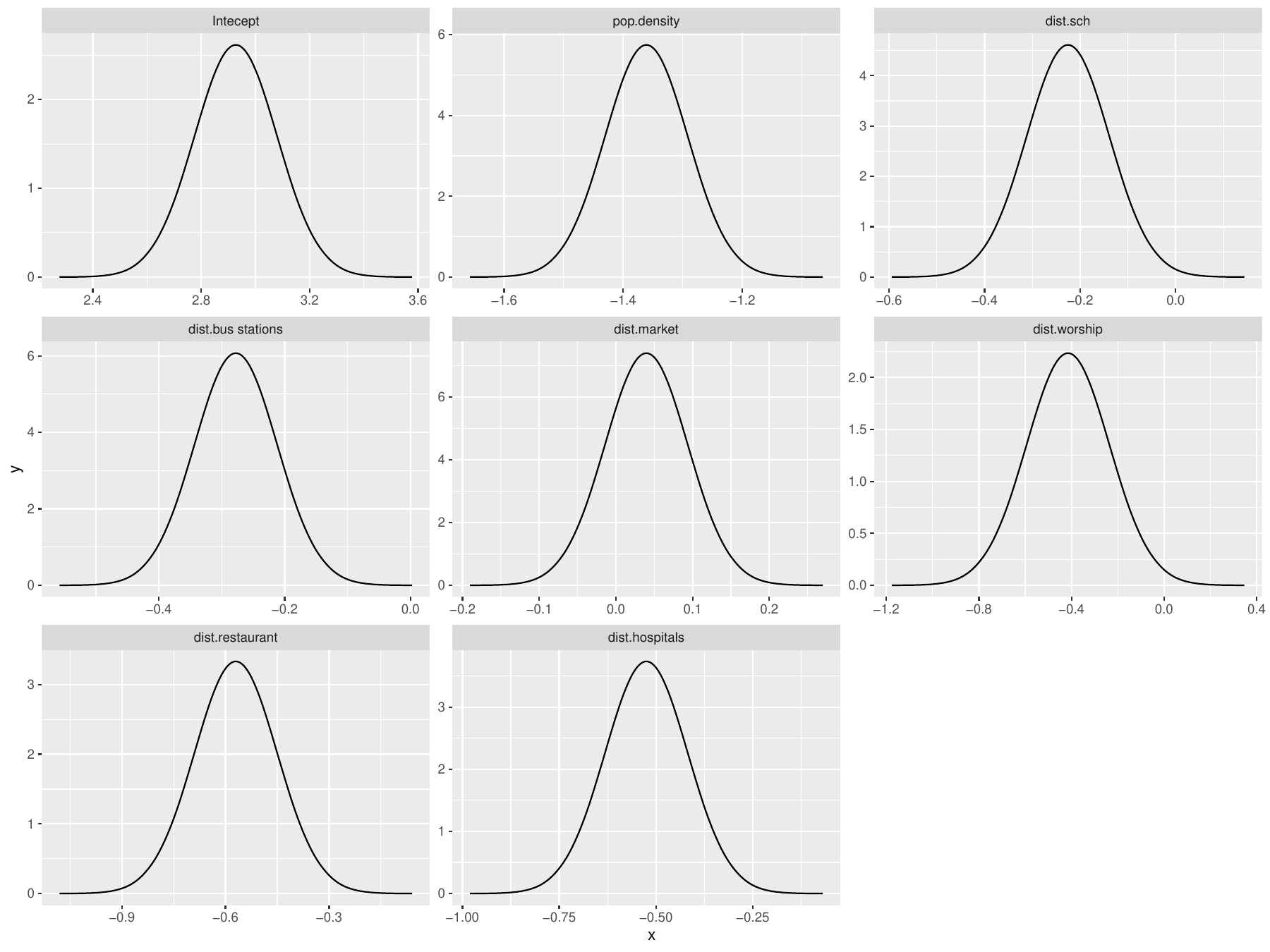}  
	\caption{Marginal posterior distributions of fixed effect coefficient from the SPDE network spatial model}
	\label{road.para}
\end{figure}

\begin{figure}[H]
	\centering
	\includegraphics[width=1\linewidth]{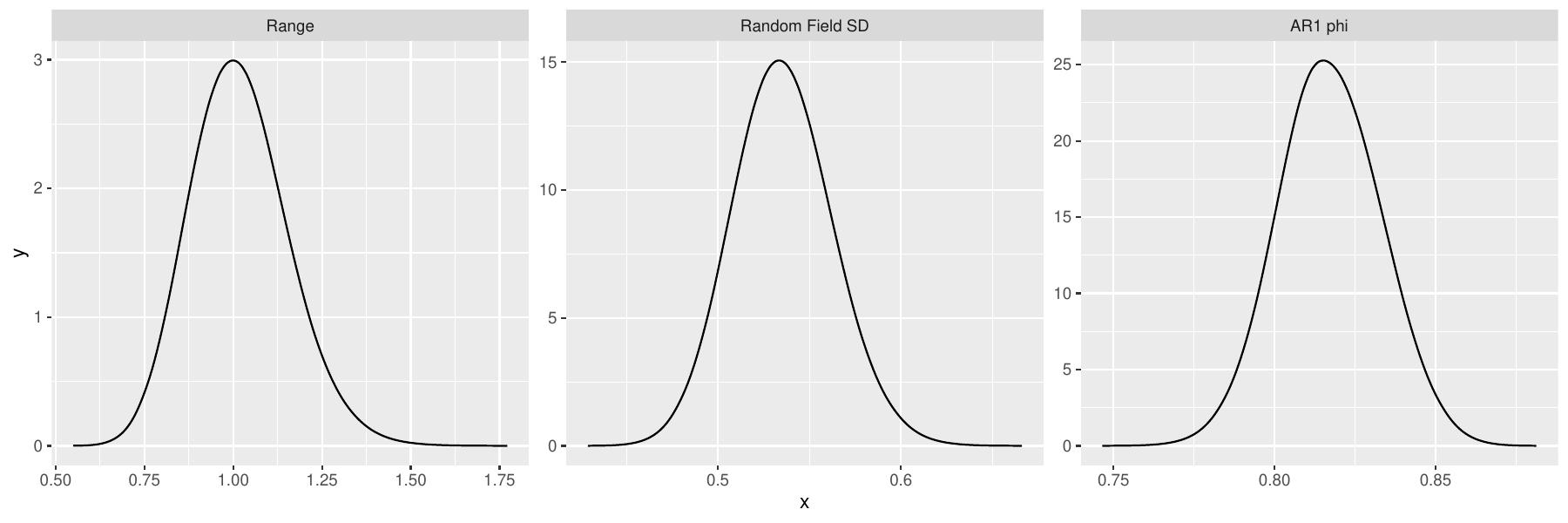}  
	\caption{Marginal posterior distributions of hyperparameters for the spatiotemporal random field of the model for the road network consideration case.}
	\label{hyper.road.para.cov}
\end{figure}

The marginal posterior distribution of the parameters for the fixed and random effects included in the model for the road network are given in Figures \ref{road.para} and \ref{hyper.road.para.cov} respectively. 
Particularly, in Figures \ref{road.para} we showed the marginal posterior distributions of all fixed effects for the road network spatial region consideration case.
Moreover, the expected values of these fixed effect parameters with corresponding 95\% credibility intervals are given in Table \ref{mean.cov.par} for the road network consideration case as well. 
In this case, the model output revealed that all the covariates except the intercept and distance from the nearest market place have a negative effect in the model like that of the entire spatial region consideration case. 
The order of the degree of negative effect, however is altered. 
The degree of the negative effect of the covariates to the model during the road network consideration case is population density, distance from nearest restaurants, distance from nearest hospitals, distance from nearest worship places, distance from nearest bus station, and distance from nearest school, in order as can be seen in Table \ref{mean.cov.par}.  
Comparing the two results of the model, the entire spatial and the road network consideration cases, the population density, and the nearest distance from restaurants are the two covariates that have a strong negative effect on the models in both cases. 
The negative effect of the covariates, nearest distance form hospitals and worship places are the third and fourth in terms of the negative influence to the model in the road network consideration case, while nearest distance from bus station and market places are the third and fourth in terms of negative effect to the model for the entire spatial region consideration case. 
The over plots of the traffic accident data on the road networks with the covarates in Figure \ref{fig:covariate-descriptive} of the Appendix II indicated that the model output for the road network consideration case is more coherent with the reality for the third and fourth covariates in terms of the negative effect to the model. This is inline with the study by \cite{okabe2012spatial}, which states traffic accident are limited to specific locations on the road networks, and considering the road networks only sounds more reasonable.
The negative effect of the population density covariate to the model in both scenarios seams contradicting if one think the more dense the population is the more the traffic fatality may become though it might not be the case always. 
For instance,  high population density could lead to traffic congestion, that could slow down vehicle speeds which potentially reduce the incident risk.
In our case, it might be because the population density obtained from WorldPop \citep{worldpop2} refers the population in the residential areas not the human traffic on the road networks for possible positive effect to the model. Moreover, it might be due to specific traffic control measures at more populous places, like the speed limits of vehicles at residential areas where the population density is high.
\begin{table}[H]  
	\caption{Marginal posterior mean and 95\% credible intervals for parameters of fixed effects for the road network models with covariates.} 
	\label{mean.cov.par}
	%\small
	\centering 
	\begin{tabular}{lccccccccccr}  
		\toprule\toprule
		\textbf{Regression parameters for covariates} &&&&&& \textbf{Mean} &&&&& \textbf{Credible intervals}\\
		\midrule[2pt] 
        Intercept ($\beta_0$) &&&&&& 2.928 &&&&& (2.629, 3.227) \\
		Population density ($\beta_1$) &&&&&& -1.361 &&&&& (-1.497, -1.225) \\
		Distance from nearest school ($\beta_2$) &&&&&& -0.225 &&&&& (-0.395, -0.056)  \\ 
		Distance from nearest bus station ($\beta_3$) &&&&&& -0.278 &&&&& (-0.406, -0.149) \\
		Distance from nearest market places ($\beta_4$) &&&&&& 0.040 &&&&& (-0.066, 0.145) \\ 
		Distance from nearest worship places ($\beta_5$) &&&&&& -0.415 &&&&& (-0.765, -0.065) \\ 
		Distance from nearest restaurant ($\beta_6$) &&&&&& -0.570 &&&&& (-0.805, -0.336)  \\ 
		Distance from nearest hospitals ($\beta_7$) &&&&&& -0.525 &&&&& (-0.734, -0.316) \\ 
		\bottomrule[2pt]  
	\end{tabular}   
\end{table}  

\begin{figure}[H]
	\centering
	\includegraphics[width=1\linewidth]{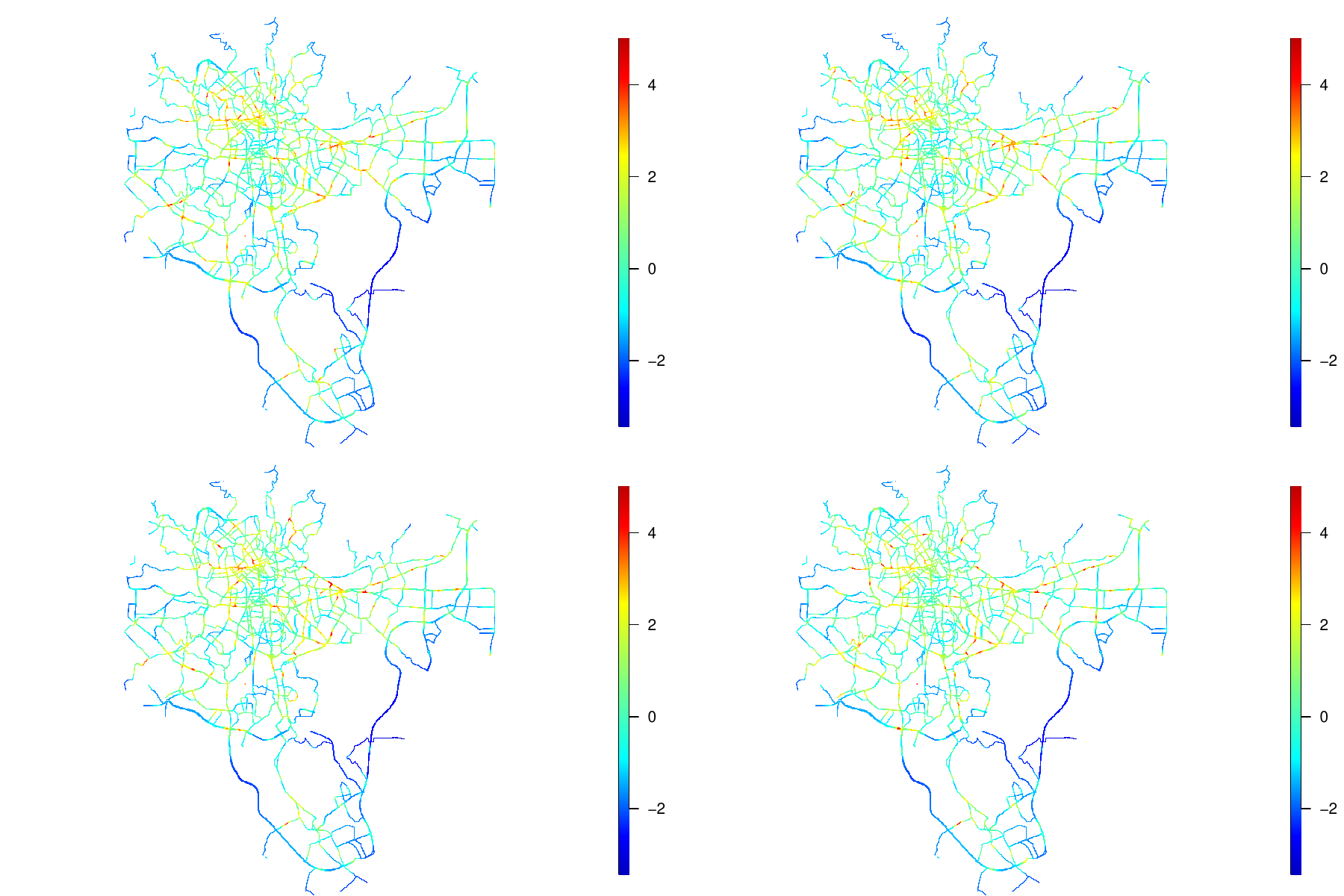}  
	\caption{Posterior mean of log-intensity surface, posterior mean for the traffic accident data on the road network: years are ordered from the top row to the bottom row, and then from left to right in each row in the sequence 2016--2019 for the model with covariate case.}
	\label{road.mod}
\end{figure}

In Figure \ref{road.mod}, we have given the log-normal mean of the intensity function estimates of the model for the road network consideration case for the study period.  
It displays a visual representation of the predicted accident intensity (in logarithmic scale) across the road network, where red shades of color indicate areas with high intensity, while lighter blue and blue colors represent areas with low intensity. 
The model fit here as well indicated how well the proposed model is able to describe the traffic accident distribution for each year on the road network during the study period 2016 -- 2019 in Addis Ababa.
The model out put showed the model's capability to identify the road segments with traffic accident hotspots for possible road safety measures by stakeholders.
The results showed a similar temporal trend of the intensity with an average temporal autocorrelation of 0.535 on the road network indicating the existence of same traffic fatality trends in the study period. 
We have given the marginal posterior distributions of the spatial effect hyperparameters: the range, standard deviation, and temporal correlation in Figure \ref{hyper.road.para.cov} for illustration purpose.
The average standard deviation of the model output for the random effect is found to be 0.817 indicating the performance of the model for such study in this particular case.

\section{Conclusions and Discussions}\label{Sec_conclusions}
In this study, we investigated the spatiotemporal dynamics of traffic accidents in Addis Ababa, Ethiopia from traffic accident data during the period 2016 to 2019.
We modelled the traffic accident as a point process and formulated a hierarchical Bayesian model with an LGCP prior assumption by considering the log-intensity as a linear combination of fixed effect and a random effect components.
The fixed effect components include covariates such as intercept, population density, distance form nearest bus station, distance from nearest schools, distance from market places, distance from restaurants, distance from worship places etc and the random effect component is considered to accommodate spatial correlations of events.
We implemented the model proposed to Addis Ababa traffic accident data spatiotemporally for the study period by considering the entire spatial region of Addis Ababa and by considering the spatial road network of Addis Ababa separately. 
The main aim of the study were to map the traffic accident dynamics spatiotemporally on the entire spatial region of Addis Ababa and on the road network of Addis Ababa and compare the effect of covariates for each domain consideration for traffic accident modeling. 
The results revealed that almost all covariates has a negative effect to the model with population density and distance from nearest restaurants has strong negative effect in order in both the entire spatial and road network consideration cases. 
For the remaining covariates, it is observed that most of them has negative effect to the model again in both scenarios though the degree of negative effect is altered for the entire spatial and road network consideration cases. The distance from nearest hospitals and worship places were third and fourth in terms of influence in the model for the road network consideration case while distance from nearest bus station and market places were third and fourth in order for the entire spatial region consideration cases. Distance from market place is seen to have negligible influence for the road network consideration case.
Comparing these results with traffic accident data and covariate locations, we conclude that the model output for the road network consideration cases outperforms the outputs for the entire spatial region consideration case. This is inline with the idea that traffic accidents are best modeled on road networks than considering off-road areas as well where accidents are unlikely to occur \citep{okabe2012spatial}.   
In general, the study outputs in both scenarios indicated the proposed model's capability in identifying traffic accident hotspots in Addis Ababa entirely and on it's road networks where traffic accidents are likely to occur. 
The findings showed that traffic accidents in Addis Ababa are affected by both spatial event correlations and the covariates considered. 
Identifying accident hotspots provides an insight for possible targeted interventions by traffic management authorities. Moreover, it is useful for pedestrians and drivers to choose safe routes in Addis Ababa.
Therefore, the findings suggested that implementing focused safety measures in the identified high-risk areas could significantly reduce accident rates. Furthermore, the study emphasized the importance of incorporating both fixed and random effects in modeling traffic accidents, providing a deeper understanding of the spatial and covariate relationships among incidents.

The model's performance with covariate and without covariate was validated using Deviance Information Criterion (DIC) and Watanabe-Akaike Information Criterion (WAIC) of statistical measures for both the entire spatial region and road network consideration cases. 
In both cases, we found minimal DIC and WAIC when the model with covariate is considered, confirming that the model captures the underlying processes that governs the traffic accidents in the study area when covariates are considered.
In conclusion, this work presented a model that can provide accurate predictions of traffic accidents that can help to give an insight for possible guided mitigation strategies. 
Moreover, the Bayesian hierarchical formulation with LGCP prior is proved to be a powerful tool to model point processes as well and analyse corresponding spatiotemporal data.

\subsection{Future Directions}
This study proved that the proposed model best performs when covariates are considered in both the entire spatial and road network SPDE mesh consideration cases of Addis Ababa.
However, the covariates considerd in this study are limited to population density and distance from nearest restaurants, from nearest bus stations, from nearest schools, from nearest hospitals, from market places and from nearest worship places only. 
We believe that it will be reasonably good if other covariates such as traffic volumes and speed limits in road segments, road types (like highway, arterial, residential), time of a day (effects of rush hours, nighttime, etc), day of weeks, weather and climatic conditions are incorporated in the model as the proposed model might be improved.
In our subsequent study, we plan to consider more covariates and extend the study to Ethiopia's road network by considering the countries major highways and regional capital's road networks. 
We are also interested to study the time series of traffic accidents to draw hotspots seasonal and trend information in detail for possible precise hotspot forecast with uncertainty.  

\paragraph{Acknowledgements:}
This work has been supported by the International Mathematical Union (IMU) and by the Graduate Research Assistantships in Developing Countries (GRAID) Program.
We would like to express our sincere appreciation to the Addis Ababa Traffic Police Office and the Addis Ababa Traffic Management Agency for kindly providing the necessary data for our research.

\bibliography{refs}
%%%%%%%%%%%%%

\appendix	
\section*{Appendix I}
\begin{figure}[H]
    \centering
    \includegraphics[width=.7\linewidth]{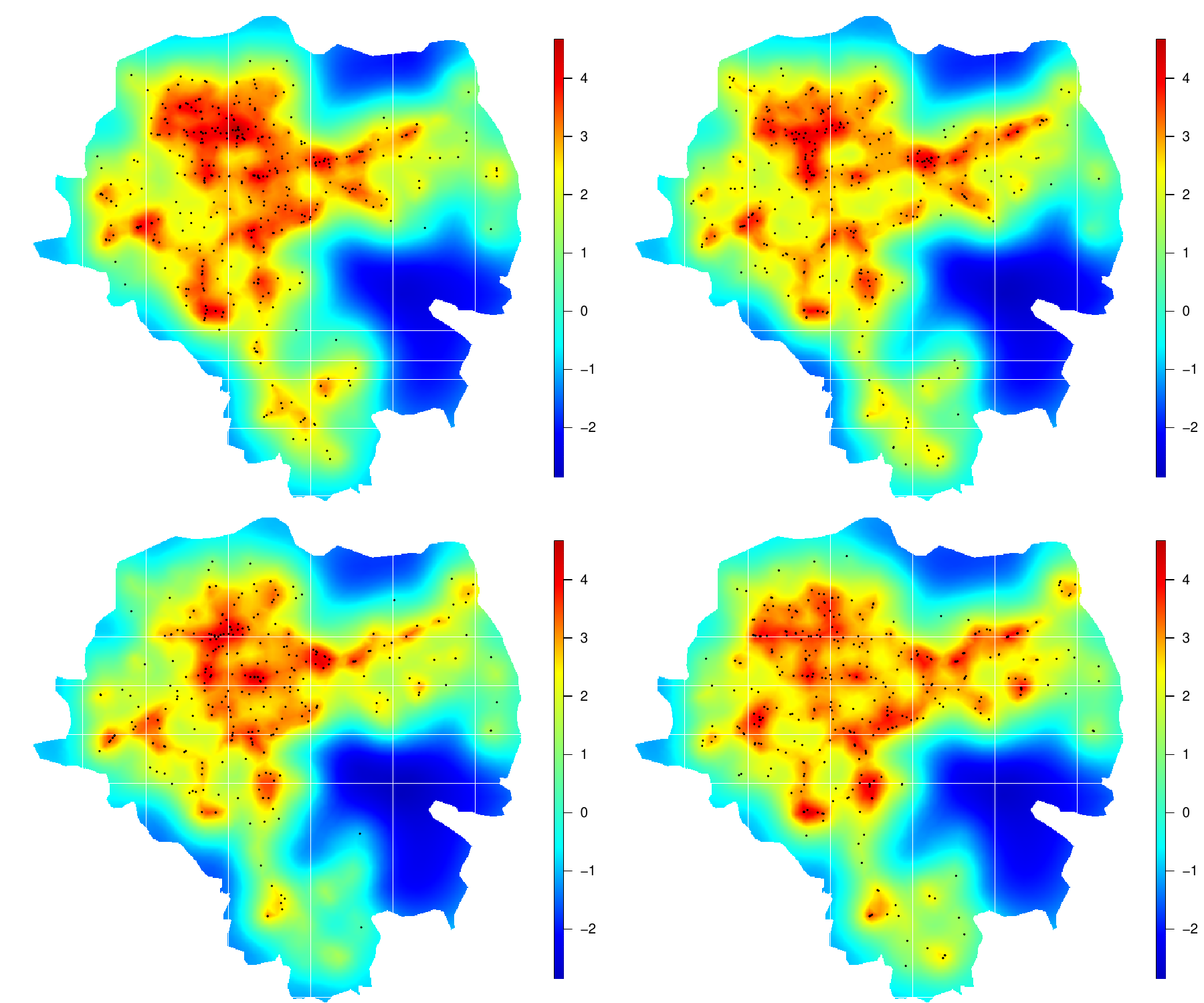}
    \caption{Maps of the log-intensity function posterior mean for the traffic accident data with out covarites on the spatial domain of Addis Ababa: years are ordered from the top row to the bottom row, and then from left to right in each row in the sequence 2016 -- 2019 }
    \label{fig:model-spatial}
\end{figure}

\begin{figure}[H]
    \centering
    \includegraphics[width=.7\linewidth]{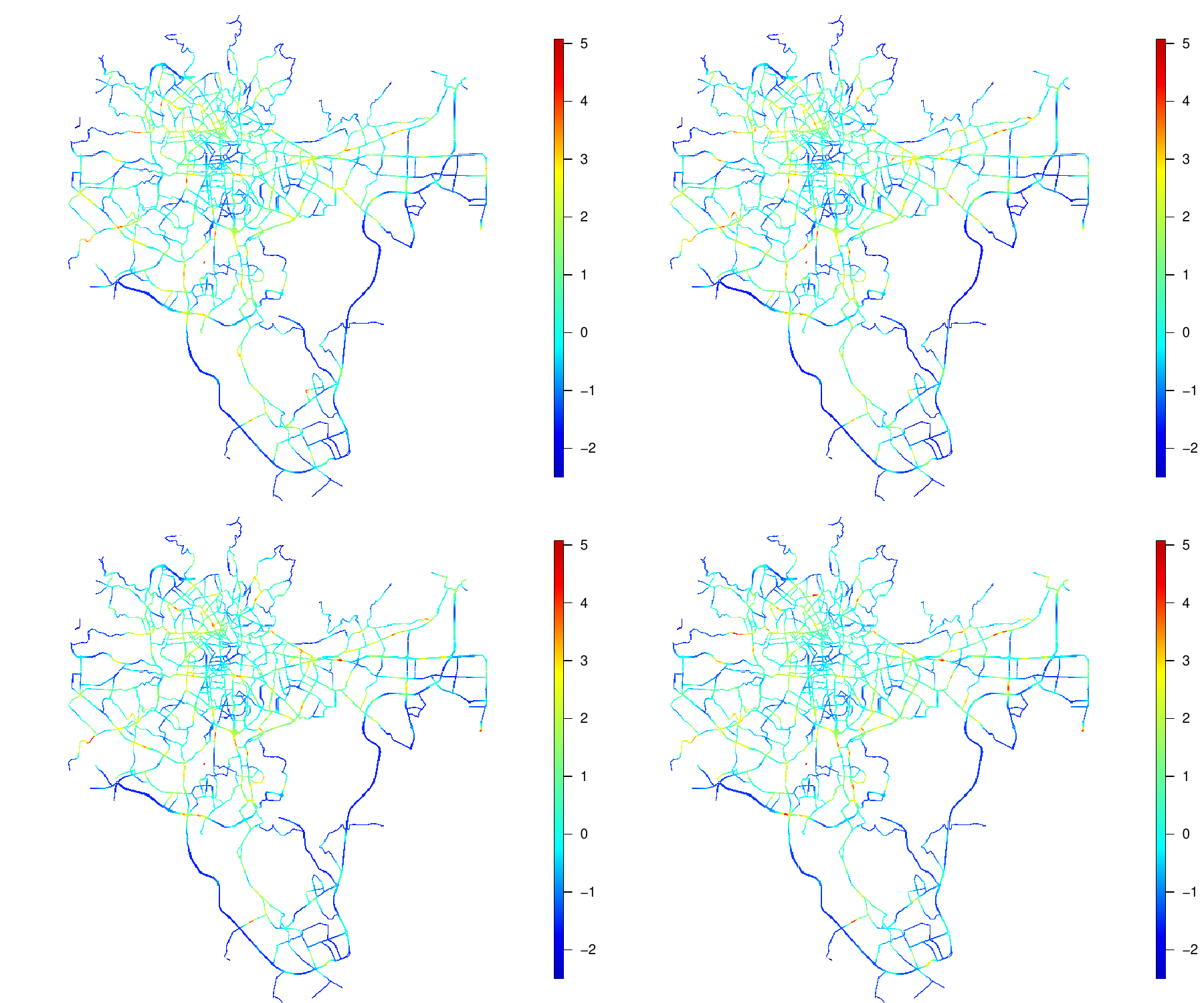}
    \caption{Maps of the log-intensity function posterior mean for the traffic accident data with out covarites on roads of Addis Ababa: years are ordered from the top row to the bottom row, and then from left to right in each row in the sequence 2016 -- 2019 }
    \label{fig:model-road}
\end{figure}

\section*{Appendix II}
\begin{figure}[h!]
    \centering
    \includegraphics[width=.9\linewidth]{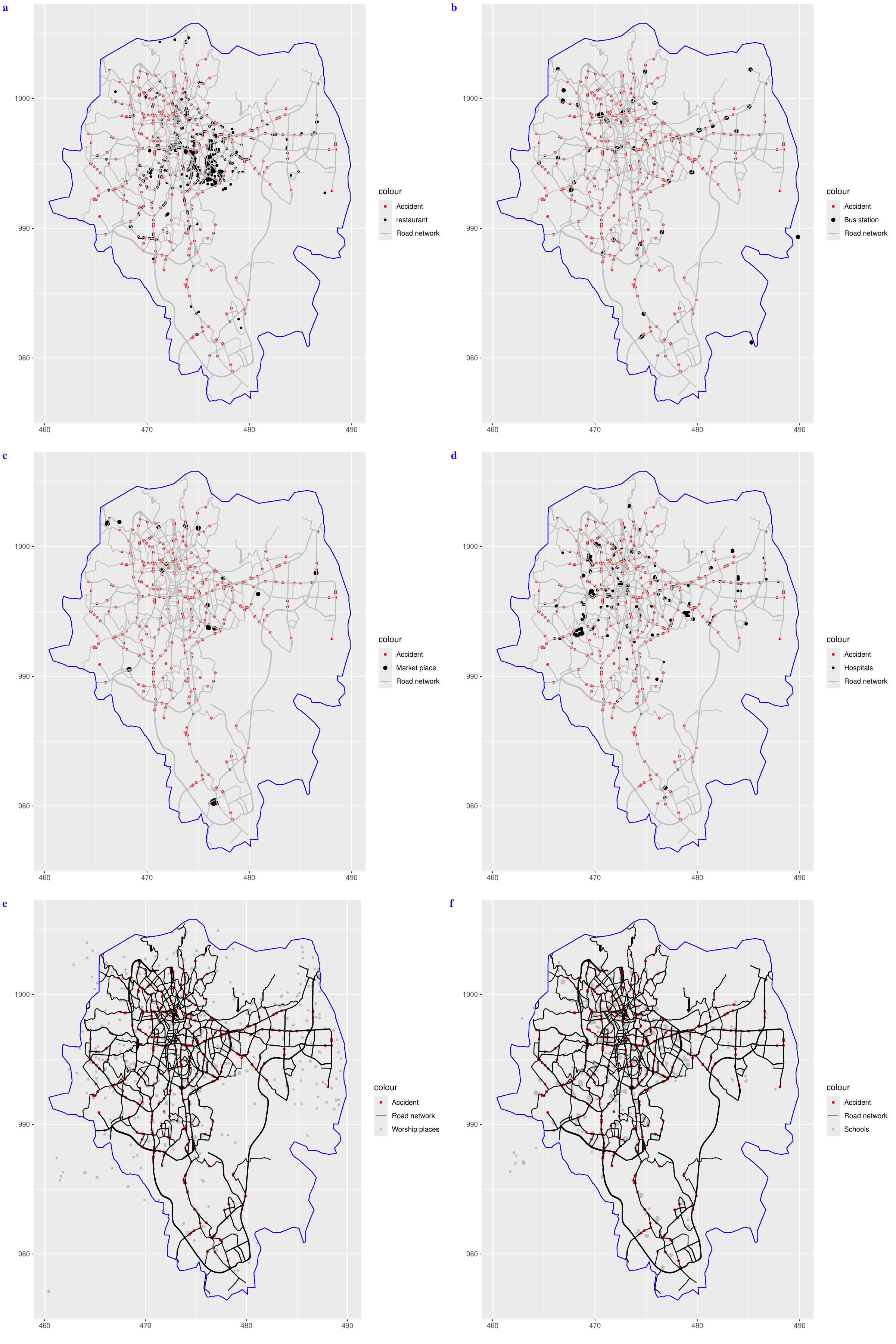}
    \caption{Locations of covariates used in the model for the year of 2016 together with the accident location (shown all in red points) and road network of Addis Ababa.}
    \label{fig:covariate-descriptive}
\end{figure}

\end{document}